\pdfoutput=1

\documentclass[11pt]{article}

\usepackage{ACL2023}

\usepackage{times}
\usepackage{latexsym}
\usepackage{listings}
\usepackage{xcolor}
\usepackage{todonotes}
\usepackage{spverbatim}

\usepackage[T1]{fontenc}

\usepackage[utf8]{inputenc}

\usepackage{microtype}
\usepackage{amsmath}
\usepackage{graphicx}
\usepackage{float}
\usepackage{subcaption}

\usepackage{amssymb}
\usepackage{blindtext}

\usepackage{booktabs}
\usepackage{caption}

\usepackage{inconsolata}
\usepackage{mathtools}
\usepackage{xcolor, soul}
\usepackage{adjustbox}
\usepackage{multirow}
\usepackage{tabularx}
\usepackage[english, ruled,vlined, boxed]{algorithm2e}
\usepackage{algpseudocode}
\usepackage{dirtytalk}

\title{Is Next Token Prediction Sufficient for GPT? Exploration on Code Logic Comprehension}

\author{
     Mengnan Qi\textsuperscript{1,*} 
     \ Yufan Huang\textsuperscript{1,*} 
     \ Yongqiang Yao\textsuperscript{1,*} 
     \ Maoquan Wang\textsuperscript{1,*}   \\
     \ \textbf{Bin Gu}\textsuperscript{2,3} 
     \ \textbf{Neel Sundaresan}\textsuperscript{1} \\
    \textsuperscript{\rm 1} Microsoft Cloud and AI \\
    \textsuperscript{\rm 2} School of Artificial Intelligence, Jilin University \\
    \textsuperscript{\rm 3} Mohamed bin Zayed University of Artificial Intelligence \\
    {\tt\{yufanhuang,mengnanqi,yongqiangyao,maoquanwang\}@microsoft.com}
 }

\begin{document}
\maketitle
\begin{abstract}

\end{abstract}

Large language models (LLMs) has experienced exponential growth, 
they demonstrate remarkable performance across various tasks.
Notwithstanding, contemporary research primarily centers on enhancing the size and quality of pretraining data, still utilizing the next token prediction task on autoregressive transformer model structure. The efficacy of this task in truly facilitating the model's comprehension of code logic remains questionable, we speculate that it still interprets code as mere text, while human emphasizes the underlying logical knowledge. In order to prove it, we introduce a new task, "Logically Equivalent Code Selection," which necessitates the selection of logically equivalent code from a candidate set, given a query code. Our experimental findings indicate that current LLMs underperform in this task, since they understand code by unordered bag of keywords. To ameliorate their performance, we propose an advanced pretraining task, "Next Token Prediction+". This task aims to modify the sentence embedding distribution of the LLM without sacrificing its generative capabilities. Our experimental results reveal that following this pretraining, both Code Llama and StarCoder, the prevalent code domain pretraining models, display significant improvements on our logically equivalent code selection task and the code completion task.

\section{Introduction}

Recent researchers in LLMs have generally followed the data-driven research paradigm of improving model performance by providing higher quality and more diverse data in combination with transformer decoder-only model architectures with large parameter sizes. Several noteworthy studies have shown outstanding performance on code-related tasks. They often adopt a strategy that input incorporating multilingual code texts, docstrings, and relevant descriptive texts, along with a huge amount of text from the other domains, into the model during the pretraining phase, which the model learned these corpus through the straightforward next token prediction task. Furthermore, there are currently specialized Code LLMs that enhance the performance and adaptability of LLMs in specific tasks within the software engineering domain through fine-tuning and additional training. Early work emphasized pretraining data and model size, GPT-C \citep{svyatkovskiy2020intellicode} is based on the GPT-2 \citep{radford2019language} model and is retrained on a large-scale unsupervised multilingual source code dataset. Expanding along this line of thought, models such as Codex \citep{chen-2021-evaluating}, GPT-NeoX \citep{black2022gpt}, have emerged. Recent work has begun to emphasize the quality of data. Phi-1 \citep{gunasekar2023textbooks} uses only a model with a size of 1.3B and is trained on carefully screened “textbook” synthetic datasets. In addition, some scholars use instruction tuning, to enhance the performance of LLMs on code data, which include but are not limited to CodeAlpaca \citep{codealpaca}, WizardCoder \citep{luo-2023-wizardcoder}, and Octopack \citep{muennighoff2023octopack}. 

\begin{figure*}[!htb]
\center{\includegraphics[width=15cm]  {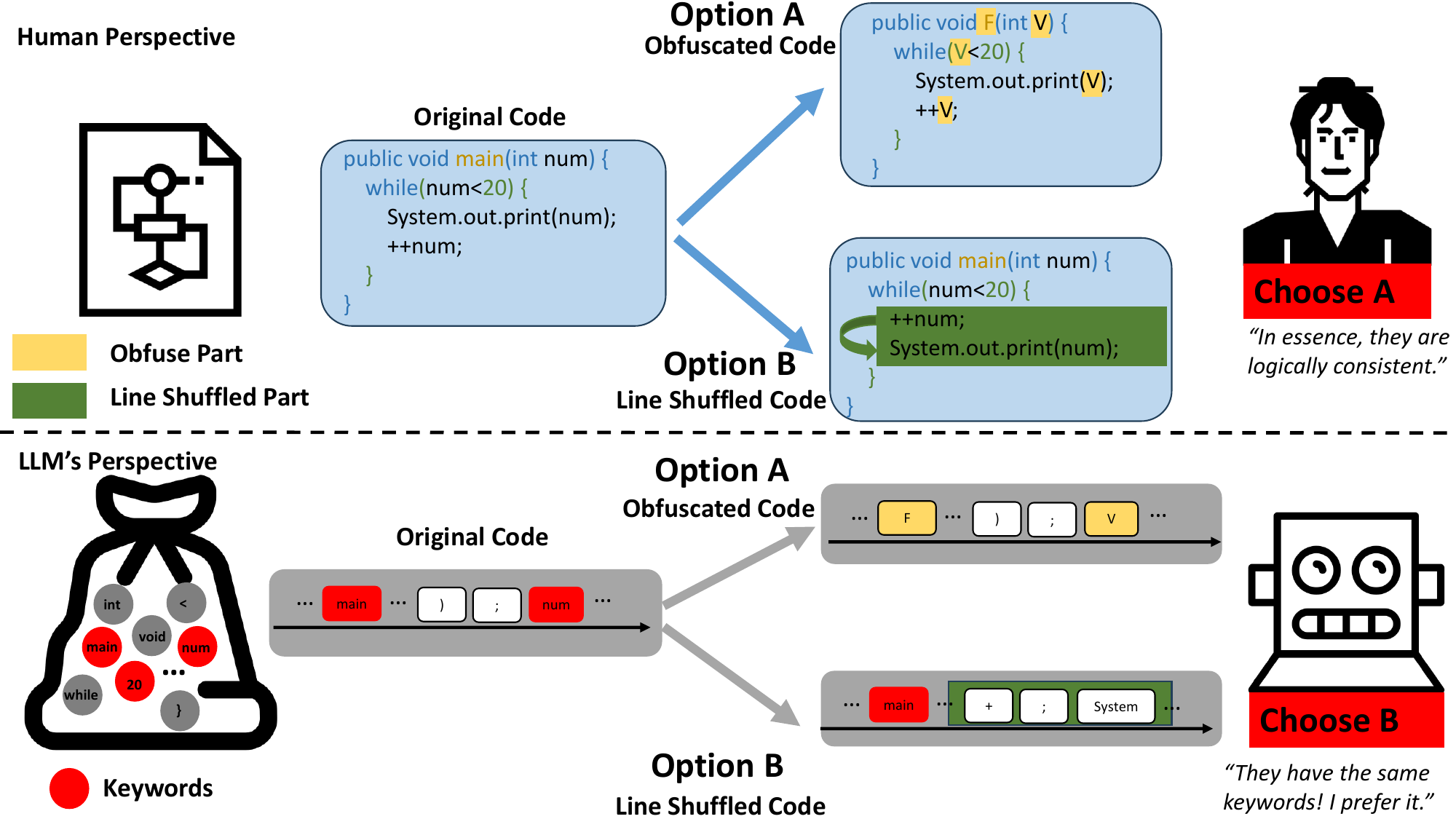}}
\caption{\textbf{The hypothesis of different perspective between human and LLM about code.} We hypothesis that humans tend to understand the logical structure behind the code (the part in the blue box) while the model tends to treat it as unordered keywords(the part in the gray box). }
\label{fig:LLM EMBEDDING motivation}
\end{figure*}

While LLMs have attained significant achievements in code generation tasks, some researchers point out that these models still cannot truly understand the code logic and execution. For instance, CodeExEcutor \citep{liu2023code} has designed a new task that compensate for the lacking ability of language models in executing code, by predicting the sequence and the intermediate status of the tracking process. Inferfix \citep{jin2023inferfix} found that the hidden errors in the code are difficult to be recognized by the model. They introduce additional static analysis for error detection, positioning and classification. CoDist \citep{huang2023program} analyzes the importance of different composition in the code in detail, and found that the model only focuses on the local information of code. 

According to the above work, we speculate that the model's interpretation of the code as unordered keywords, a perspective that diverges from the human understanding that more underscores the importance of cognizance of the underlying logic. We elaborate on our hypotheses using the example in Figure \ref{fig:LLM EMBEDDING motivation}, powerful LLMs make mistakes when selecting the option that is logical equivalent to the original code. Specifically, Option A replaces the names of identifiers in the code with abstract, meaningless characters, but the logic remains unaffected. Conversely, the code logic in option B has changed as we shuffled some lines, but the keywords did not change, so the model still prefers it. This phenomenon shows the different perspective in code understanding between humans and these models.

To further verify our hypothesis, we propose a novel task, Logically Equivalent Code Selection. This task necessitates the model to discern between two candidate codes and select the one that is logically equivalent to the query code, providing insights into the model's ability in comprehending the code logic. We found that powerful LLMs performed poorly on our new tasks, this may come from the inherent limitations of the next token prediction task. In order to improve their performance, we also introduced an innovative pretraining task, Next Token Prediction+, which follows the training format of the Next Token Prediction task. In this task, we construct obfuscated code and line-shuffled code for the original code as positive samples and negative samples respectively. The original code continues to perform the next token prediction task, while the prediction sequences of the remaining two are additionally designed. This training method modifies the sentence embedding distribution of the LLM without compromising its generative abilities. We demonstrate that our approach achieves universal improvement on our logic aware task, with average improvements of 22.95\%, 23.23\% and 23.99\% on Code Llama 7b, Code Llama 13b and StarCoder 15b respectively.

Our contribution can be summarized as follows:

\begin{itemize}
    \item We found that LLMs trained on the next token prediction task currently struggle to truly understand the underlying logic of the code. Specifically, LLMs deviate from human intuition when dealing with functionally equivalent but textually diverse codes, and marginally different codes that induce significant bugs.
    \item To measure this discrepancy, we introduce a novel challenging task called Logically Equivalent Code Selection that select logically equivalent codes from the candidate set given a query code. Compared to other code comprehension tasks, this task is more capable of delving deeper into the model's understanding of code logic.
    \item To improve the performance of LLMs on the logical equivalent code selection task, we also introduce an innovative pretraining task, Next Token Prediction+, which follows the training format of the Next Token Prediction task. Our training method adjusts the sentence embedding distribution of the LLM without compromising its generative abilities. 
    \item  Our experiments show that after this pretraining, both Code Llama and StarCoder, two popular code domain pretraining models, exhibit marked improvement on our logically equivalent code selection task and the code completion task.    
\end{itemize}

\section{Related Works}

\subsection{Code Evaluation Task}
The CodeXGLUE \citep{lu2021codexglue} benchmark is the most comprehensive and influential list in the field of code evaluation. It comprises 14 datasets for 10 diversified programming language tasks. For code generation tasks, including the line-level and token-level code complement, code repair, type prediction, text-to-code generation, code summarization, cloze test which predict the masked code from several candidates and code-to-code translation which cross the different programming languages. For code understand tasks, including clone detection which retrieve semantically similar codes given a code as the query, defect detection which determine whether the given function is vulnerable and code search which select the most related code to natural language. \citep{guo2022unixcoder} introduces a zero-shot code-to-code search task. In this task, a source language code is given as the query and then model should retrieve code that share the same semantics from a pool of target language candidate codes in a zero-shot setting.

\subsection{Self-supervised Pretrain Tasks for Code}
Inspired by success of pre-trained models in natural languages (NL), pre-trained models in programming languages (PL) has been transfered to code-related tasks. CuBERT \citep{kanade2019pre} makes first attempt at pretraining the mask language model (MLM) task \citep{devlin2018bert} on a Python corpus for code-specific contextual embeddings. CodeBERT \citep{feng2020codebert} adapts replaced token detection task(RTD) to pre-train BERT on NL-PL pairs.

And there are also some encoder-decoder pre-trained models. DOBF \citep{roziere2021dobf} first uses a de-obfuscation objective to pre-train model to learn programming syntax and shows code translation can benefit from de-obfuscation task. PLBART \citep{ahmad2021unified} is based on the BART \citep{lewis2019bart} and pre-trained on NL and PL corpus using denoising objectives. CodeT5 \citep{wang2021codet5} adapts the T5 \citep{raffel2019exploring} model that considers the crucial token type information from identifiers and allow for multi-task learning on downstream tasks.

Recent large language models mostly follow the GPT-style \citep{floridi2020gpt} training format, which performs causal language model tasks in a decoder-only transformer structure. However, this format encounters difficulties when tasked with filling variable-length whitespace. An autoregressive blank infilling task was developed and introduced in the General Language Model (GLM) \citep{du2021glm}, whilst researchers proposed another solution known as the Fill in the Middle (FIM) task \citep{bavarian2022efficient}. These innovative techniques have consequently led to a marked improvement in the performance of infilling tasks.

\section{Logically Equivalent Code Selection}

In this section, we mainly introduce a new task, Logically Equivalent Code Selection, that can measure the model's ability to understand the underlying code logic. We will first provide a detailed description of how to build the task, including task format description, data construction, etc. Then we evaluate the performance of some popular large language models on this task.
\subsection{Task Description}
Our evaluation task begins with providing the query part that is a function-level logically correct code, referred to as original code ($C_{Origin}$). Subsequently, a pair of candidate codes are presented as two candidate options. The first is a logically equivalent version of the original code, deemed accepted code ($C_{Accepted}$), while the second is a logically non-equivalent variant, termed rejected code ($C_{Rejected}$). The model is then tasked with selecting one from these two options. The overall accuracy is computed based on the model's performance across all such pairs of problems. 

\subsection{Candidate Data Construction}
In this section, we describe candidate codes, including their types, form expressions, and specific construction processes.
\subsubsection{Accepted Candidate}
The accepted candidate's code will be logically equal to the code used as a query. 

\noindent \textbf{Positive Code} This part of the code is sourced from code problem solving datasets. As programming websites often provide different solutions to the same problem, we take another solution that originate from the same problem as the positive code, which form as $C_{Positive}$.

\noindent \textbf{Obfuscated Code} In this approach, we propose to apply semantic-preserving perturbation by replacing the identifiers in code snippets, which is proposed in DOBF~\cite{roziere2021dobf}. In this disturbance, variable names and function names in the code are replaced with abstract characters without semantic information. Obfuscated code deviates significantly from the original code in terms of characters, but their underlying logic remains consistent, making it a challenging positive sample. It can be expressed as $C_{Obfuscated}$.

\subsubsection{Rejected Candidate}
The code in the rejected candidate is code that is not logically equivalent or even has bugs.

\noindent \textbf{Negative Code} Corresponding to positive code, negative code is the solution to other questions, representing any normal trivial code that differs from query logic, which form as $C_{Negative}$.

\noindent \textbf{Line Shuffled Code} We propose to randomly shuffle the original code to see whether the models are sensitive to the order of tokens, which is actually represented by the positional embedding fed to the models. We randomly select some consecutive lines in the code and shuffle their order, which is a relatively subtle way to add perturbations. We ensure that the logic of the shuffled code changes or directly becomes code with serious bugs. It can be expressed as $C_{Line\_Shuffled}$.
\subsubsection{Candidate Code Pair}
We need to randomly select a type from the accepted candidate as the correct answer and a type from the rejected candidate as the incorrect answer. So we finally designed three challenging pairs of candidate data. Namely, they are $(C_{Positive},C_{Negative})$, $(C_{Obfuscated},C_{Negative})$, $(C_{Positive},C_{Line\_Shuffled})$, in which the correct answer are $C_{Positive}$, $C_{Obfuscated}$, $C_{Positive}$. We separately calculate the proportion of correct responses from the model as the evaluation metric.

\subsection{Model Performance}
In this section, we test models on our task by comparing the distance between the origin code and two candidate codes in the embedding space. We apply average pooling to the outputs from the last layer of the pre-trained model to get their embeddings and compute the cosine similarities of the embeddings to represent the distance to the origin code, like $D(Positive) = cos(E_{Origin}, E_{Positive})$. So the corresponding metric can be formalized like $Acc(D(Positive)>D(Negative))$. We test three candidate code pairs which proposed in the Section 3.2.3 and conduct experiments on the Code Llama and StarCoder models.

\begin{figure}[!htb]
\center{\includegraphics[width=7.5cm]  {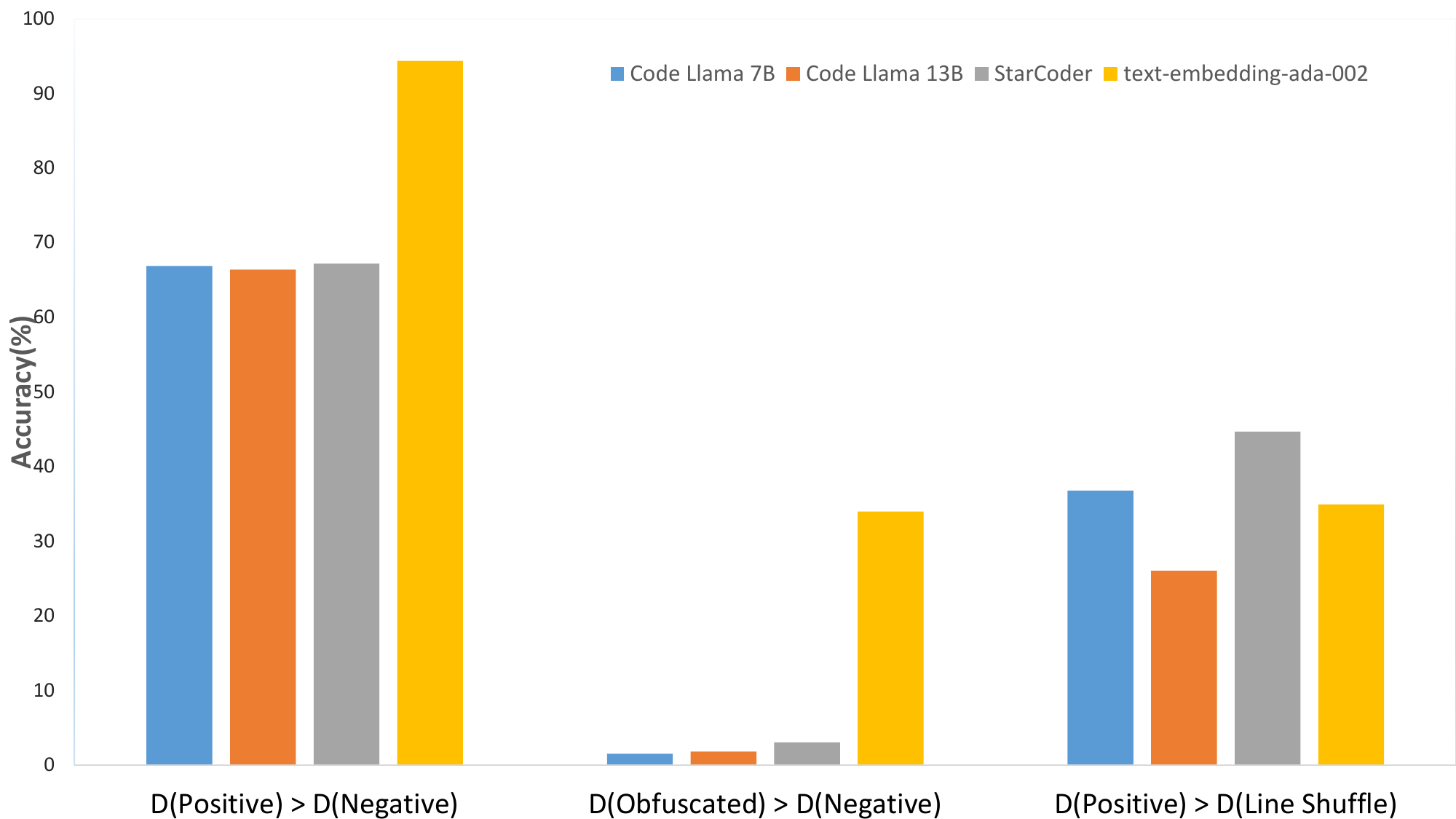}}
\caption{\textbf{The accuracy distribution of different models on different embedding similarity analysis subtasks.} The blue, orange, and gray columns respectively represent the accuracy of large language models of different sizes (7B, 13B, 15B) on the six distance analysis tasks we have designed. Additionally, we have included OpenAI's specialized embedding model, text-embedding-ada-002 (depicted by a yellow column), as a control for comparison.}
\label{fig:LLM embedding overarching framework}
\end{figure}

As shown in Figure \ref{fig:LLM embedding overarching framework}, the obfuscated code demonstrates a larger divergence from the original code than the negative code across all examined models. Furthermore, it is discernible that Code Llama and Starcoder models of different versions exhibit a deviation from human intuition in their comprehension of obfuscated code and line-shuffled code. Despite the obfuscated code wholly preserving the logic of the original code, the model perceives it as less relevant to the original code compared to a randomly selected negative code. This is attributed to the obfuscation of key tokens that encapsulate the code semantics. In contrast, the line-shuffled code, which disrupts the structure of the code, represents the least conspicuous perturbation. Paradoxically, the model perceives its correlation to exceed that of the positive code when compared with other perturbation schemes. In order to comprehensively explore the key points of the model's understanding of code text, we also analyzed the performance of other forms of perturbation in Appendix \ref{sec:Extra_Perturbations_Analysis}. These findings suggest that when a large model analyses code, it predominantly relies on unordered key tokens, failing to comprehend the underlying logical information.

\section{Method}
In this section, we describe how to improve the performance of large language models using a new self-supervised pretraining task in the continuous pretraining stage. Our solution adjusts the prediction targets of hard negative samples and hard positive samples based on the next token prediction task form. It is worth noting our negative sample training part, which is an innovative method that forces the model not to learn wrong code forms. 

Prominent large language models currently in use frequently adopt the model structure and pretraining techniques of GPT. This is due to the ease of model scaling and the reduction of discrepancies between pretraining and downstream fine-tuning. As our method is applied to continuous pretraining stage, it is better to align with the pretraining tasks employed by these models, the next token prediction task, which is optimally suited for auto-regressive models.

\noindent \textbf{Original Code. } In this section, we precisely align the process with the pre-existing pre-trained model. We filter out high-quality, method-level code data from the pretraining code corpus to ensure compatibility with syntactic parsing tools. We then subject this data to the next token prediction task. Given an unsupervised code of tokens $C_{origin} = \{c_{0}^{ori},...,c_{n}^{ori}\}$, the loss can be defined as follows:

\begin{equation} 
L_{ori}(\theta) = \sum_{i} \log P \left(c_i^{ori} | c_0^{ori}, \ldots, c_{i-1}^{ori}; \theta \right)
\end{equation} 

\noindent \textbf{Line Shuffled Code.} As hard negative samples, the objective of incorporating noise into the original code is to create a version that, on the surface, closely mirrors the original code but harbors divergent logic or even potential bugs. To facilitate this design, we attempt to choose several consecutive lines of code from within the method and randomly invert their sequence. It is crucial to note that certain statements within the code may lack logical dependencies, implying that even if their order is reversed, the code maintains logical equivalence. We employ dependency analysis tools to identify and filter out such instances.

\begin{figure}[!htb]
\center{\includegraphics[width=7.5cm]  {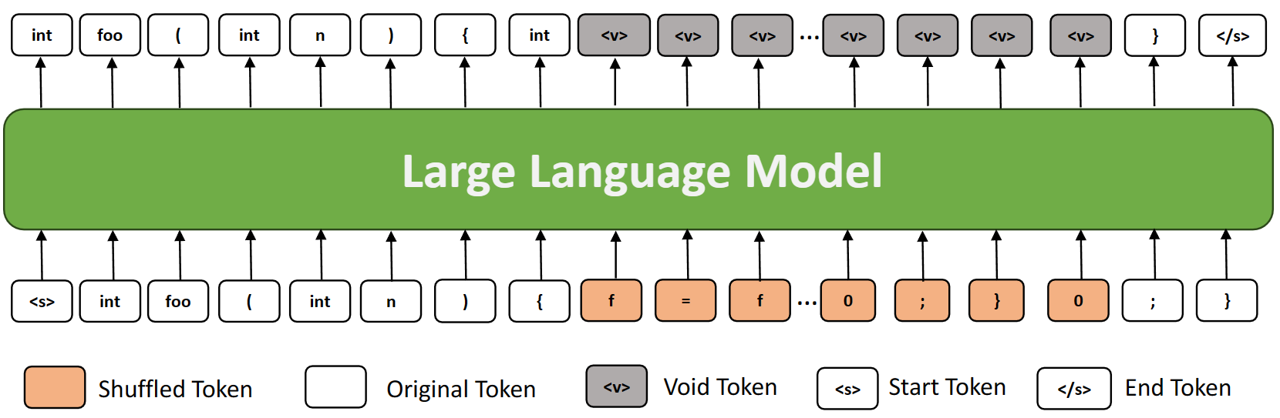}}
\caption{\textbf{The next token prediction task on line shuffled  code.} In the illustrative diagram, the white box symbolizes the normal token, while the orange box represents the shuffled token. The intermediate output is pointed towards a uniquely designed "void" token <v>. This methodology enables the model to bypass the problematic portions of the code, whilst retaining its comprehension of the normal sections.}
\label{fig:negative pretrain}
\end{figure}

Distinct from the aforementioned scenario, line-shuffled code is frequently not standard code, necessitating the prevention of the model from learning this erroneous form. To achieve this, we compel the model to generate new sequences wherein the portion of tokens that has been injected with noise is directed towards a unique "void" token, whereas the normal part is trained to predict the subsequent token. This strategy ensures that the model does not learn incorrect coding practices from the noisy data. Consider the line shuffled code as $C_{line\_shuffled} = \{c_{0}^{lsf},...,c_{n}^{lsf}\}$, in which the $j<=i<=k$ part is the scrambled code, then the output should be $Y = \{y_{1},...,y_{n+1}\}$. Training details are shown in Figure \ref{fig:negative pretrain}, and the loss can be defined as follows:

\begin{equation} 
L_{lsf}(\theta) = \sum_{i} \log P \left(y_i | c_0^{lsf}, \ldots, c_{i-1}^{lsf}; \theta \right)
\end{equation} 

\noindent \textbf{Obfuscated Code.} As hard positive samples, the objective of designing obfuscated code is to generate code that is functionally identical, yet exhibits maximal textual variation. This methodology aligns with the approach used in DOBF~\cite{roziere2021dobf}, whereby variable names and function names that embody substantial semantic information are renamed to more abstract identifiers such as Vi and Fi. Throughout this renaming process, we maintain that both the declaration and all subsequent usages of any renamed variable are updated consistently. Moreover, we ensure that a unique identifier is not recycled for the renaming of more than one variable.

\begin{figure}[!htb]
\center{\includegraphics[width=7.5cm]  {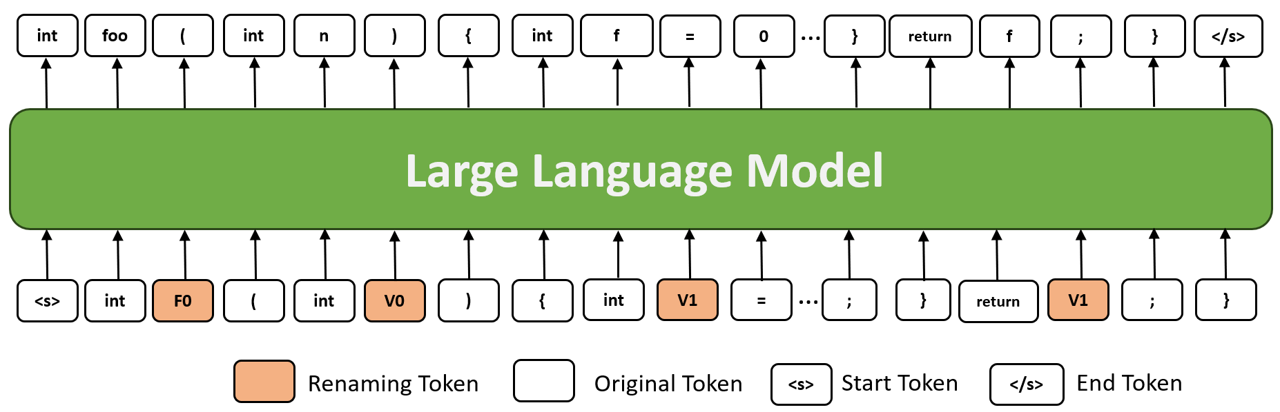}}
\caption{\textbf{The next token prediction task on obfuscated code.} This task continues to adhere to the training format of the causal language model. In the depicted diagram, the white boxes denote normal tokens, while the orange boxes represent obfuscated tokens. Throughout the training process, the obfuscated code acquires the output embedding via the model, and attempts to optimize the loss with the target token through the linear layer. This technique aims to diminish the output distribution gap between the obfuscated code and the corresponding normal code.}
\label{fig:positive pretrain}
\end{figure}

We expect that the obfuscated code can be as close as possible to its corresponding original code containing the actual key information token in the output distribution. We use the obfuscated code as input and the corresponding original code as the label sequence. Suppose the obfuscated code of $C_{obfuscated} = \{c_{0}^{obf},...,c_{n}^{obf}\}$. Training details are shown in Figure \ref{fig:positive pretrain}, and the loss the loss can be defined as follows:

\begin{equation} 
L_{obf}(\theta) = \sum_{i} \log P \left(c_i^{ori} | c_0^{obf}, \ldots, c_{i-1}^{obf}; \theta \right)
\end{equation}

In pretrain stage, our objective is to leverage self-supervised pretraining to glean enhanced pre-trained representations from unlabeled data. When presented with an input of method-level code, we procure its corresponding positive and negative samples. These samples are then input into the same model. We optimize these three objectives concurrently, culminating in the final loss, which is expressed as follows:

\begin{equation} 
L = L_{ori} + L_{obf} + L_{lsf}
\end{equation} 


\section{Experiments}
In this section, we will explain our experimental settings and report the results. To evaluate our approach and show its universal impact, we conduct experiments on 3 different tasks, including a code understanding task and two code generation task.
\subsection{Pretrain Details}
\textbf{Training Datasets}
We select Java and Python files from projects with more than 5 stars in the GitHub public repositories. We extract complete, structurally sound code at the function level, and further filter out high-quality code through deduplication, syntax parser, length control, standard formatting.

\noindent \textbf{Experimental Details}
For continual pretraining of large language models, we choose StarCoder and Code Llama for experiment. In the handling of obfuscated codes, our strategy for augmenting data diversity involves randomly selecting a portion of the variable names and function names to maintain the original semantics during code obfuscation. On the other hand, for line shuffled codes, 
we randomly select the code line numbers at the beginning and end of the part that needs to be shuffled and use a parser generator tool called TreeSitter\footnote{\url{https://tree-sitter.github.io/tree-sitter/}} to select buggy code as training data. More training details are shown in Appendix \ref{sec:training_details}.
\subsection{Code Understanding Task}
Firstly, we examine the performance of the continuously pre-trained large language models on our novel task, the Code Logic Equivalence Determination Task. We build our test set starting from the Project CodeNet Java250 dataset, which is a subset of the Java language in CodeNet~\cite{puri2021codenet}, a comprehensive dataset accumulated from various online programming platform. This subset encompasses 1,400 questions expressed in Java code, each accompanied by multiple solution variants. We randomly extract 16,326 triple code pairs (original, positive, negative) from this subset, and simultaneously build obfuscated code and line shuffled code from the original code.

Table \ref{table:Code_Logic_Equivalence_Determination_Task_Results} presents the accuracy of Code Llama (7B, 13B) and StarCoder on three subtasks. The first column reflects the model's capacity to provide a reasonable embedding representation distribution. The specific judgement criterion is the occurrence ratio of D(Positive)>D(Negative). The second column depicts the model's ability to distinguish codes that, despite having the greatest textual semantic difference from the original code, are logically equivalent and any trivial negative code. The specific judgement criterion is the occurrence ratio of D(Obfuscated)>D(Negative). The third column outlines the model's ability to distinguish logically inequivalent codes from the original code and any ordinary positive code, with subtle perturbations in the text dimension. The specific judgement criterion is the occurrence ratio of D(Postive)>D(Line Shuffled).

\captionsetup[table]{font={small}}
\begin{table}[ht]

\begin{adjustbox}{width=\linewidth}
\begin{tabular}{l|ccc}
\hline
\textbf{Model} & $D(Pos)>D(Neg)$ & $D(Obf)>D(Neg)$ & $D(Pos)>D(LSh)$ \\
\hline
Code Llama 7B & {66.91\%} & {1.54\%} & {36.82\%} \\
+ Pretrain & {79.6\%} & {35.13\%} & {59.40\%} \\
Code Llama 13B & {66.43\%} & {1.83\%} & {26.08\%} \\
+ Pretrain & {77.96\%} & {35.54\%} & {50.54\%} \\
StarCoder 15B & {67.19\%} & {3.08\%} & {44.72\%} \\
+ Pretrain & {80.48\%} & \textbf{38.11\%} & \textbf{68.36\%} \\
\hline
text-embedding-ada-002 & \textbf{94.4\%} & {33.97\%} & {34.95\%} \\

\bottomrule
\end{tabular}
\end{adjustbox}

\caption{\raggedright {\textbf{Compare the model performance in Code Logic Equivalence Determination Task.} } }
\label{table:Code_Logic_Equivalence_Determination_Task_Results}

\end{table}

From the results, we observe that post our pretraining, the accuracy of the three tested models in all the three subtasks has been significantly enhanced, indicating that the embedding distribution of the model has also become more reasonable.

\subsection{Code Generation Task}

In addition to enhancing the code representation capabilities of large language models, it is also our aim to preserve their impressive generative capabilities. To gauge this, we have selected the code completion task from the CodeXGlue benchmark which contains the PY150 and Github Java Corpus datasets and assessed it at two granular levels: the token level and the line level.

\noindent \textbf{Token-level code completion} involves predicting the next token (right side) given a previous code token as context (left side). we employ the token generation accuracy (Acc) as metric, the results are shown in Table \ref{table:Code_Completion_Task_Results_for_token}.

\noindent \textbf{Line-level code completion} resembles token-level prediction, however, the objective of the model here is to predict the next token until the entire line of code is completed (i.e., not merely predicting one next token). For line-level predictions, we utilize the same model employed for token-level code completion tasks to iteratively generate the next token. The newly produced token is used as context for the subsequent prediction. This procedure is repeated iteratively until the model generates an <EOL> tag. 

we employ exact match (EM) and edit similarity (ES) as metric and show the results in Table \ref{table:Code_Completion_Task_Results_for_line}. The parameter of exact match (EM) necessitates a perfect congruence between the predicted line of code and the ground truth line, allowing for no deviations. On the other hand, edit similarity (ES) quantifies the minimum number of character edits, encompassing insertion, deletion, or substitution of characters, that are needed to match the predicted row to the actual row.

\captionsetup[table]{font={small}}
\begin{table}[ht]

\begin{adjustbox}{width=\linewidth}
\begin{tabular}{l|cc}  
\hline  
\textbf{Model} & PY150 & Java Corpus \\   
\hline
Code Llama 7B & {83.57\%} & {83.11\%} \\
+pretrain & {86.51\%} & {86.69\%} \\
Code Llama 13B & {87.11\%} & {86.42\%} \\
+pretrain & \textbf{89.86\%} & \textbf{88.05\%} \\
StarCoder 15B & {86.9\%} & {83.17\%} \\
+pretrain & {87.23\%} & {84.49\%} \\
\hline 
CodeGPT-adapted & {76.60\%} & {77.73\%} \\
PyCoder & {76.93\%} & {-} \\

\bottomrule
\end{tabular}
\end{adjustbox}

\caption{\raggedright {\textbf{Compare the model performance in Token-level code completion Task.}  } }
\label{table:Code_Completion_Task_Results_for_token}

\end{table}





\captionsetup[table]{font={small}}
\begin{table}[ht]

\begin{adjustbox}{width=\linewidth}
\begin{tabular}{l|cccc}  
\hline  
\textbf{Model} & \multicolumn{2}{c}{PY150} & \multicolumn{2}{c}{Java Corpus} \\  
\hline 
Metric & EM & ES & EM & ES \\
\hline
Code Llama 7B & 50.79 & 75.26 & 40.07 & 73.21 \\
+pretrain & 52.14 & 77.31 & 42.61 & 75.53 \\
Code Llama 13B & 54.28 & 77.64 & 42.11 & 74.27 \\
+pretrain & \textbf{57.16} & 80.95 & \textbf{45.22} & \textbf{76.38} \\
StarCoder 15B & 54.27 & 78.17 & 40.18 & 74.22 \\
+pretrain & 56.13 & \textbf{81.16} & 42.05 & 73.93\\
\hline 
CodeGPT-adapted & 42.37 & 71.59 & 30.60 & 63.45 \\
CodeT5+ 770M & 44.86 & 74.22 & 37.90 & 72.25 \\
PyCoder & 43.91 & 71.74 & - & - \\

\bottomrule
\end{tabular}
\end{adjustbox}

\caption{\raggedright {\textbf{Compare the model performance in Line-level code completion Task.}  } }
\label{table:Code_Completion_Task_Results_for_line}

\end{table}






It is noteworthy that the generative capacity of the pre-trained model remains undiminished. In fact, it exhibits quicker convergence following the same fine-tuning process, thereby leading to an overall enhancement in performance.

\section{Analysis}
\subsection{Similarity Visualization}

\begin{figure*}
  \centering
  \begin{minipage}[b]{0.45\textwidth}
    \includegraphics[width=\textwidth]{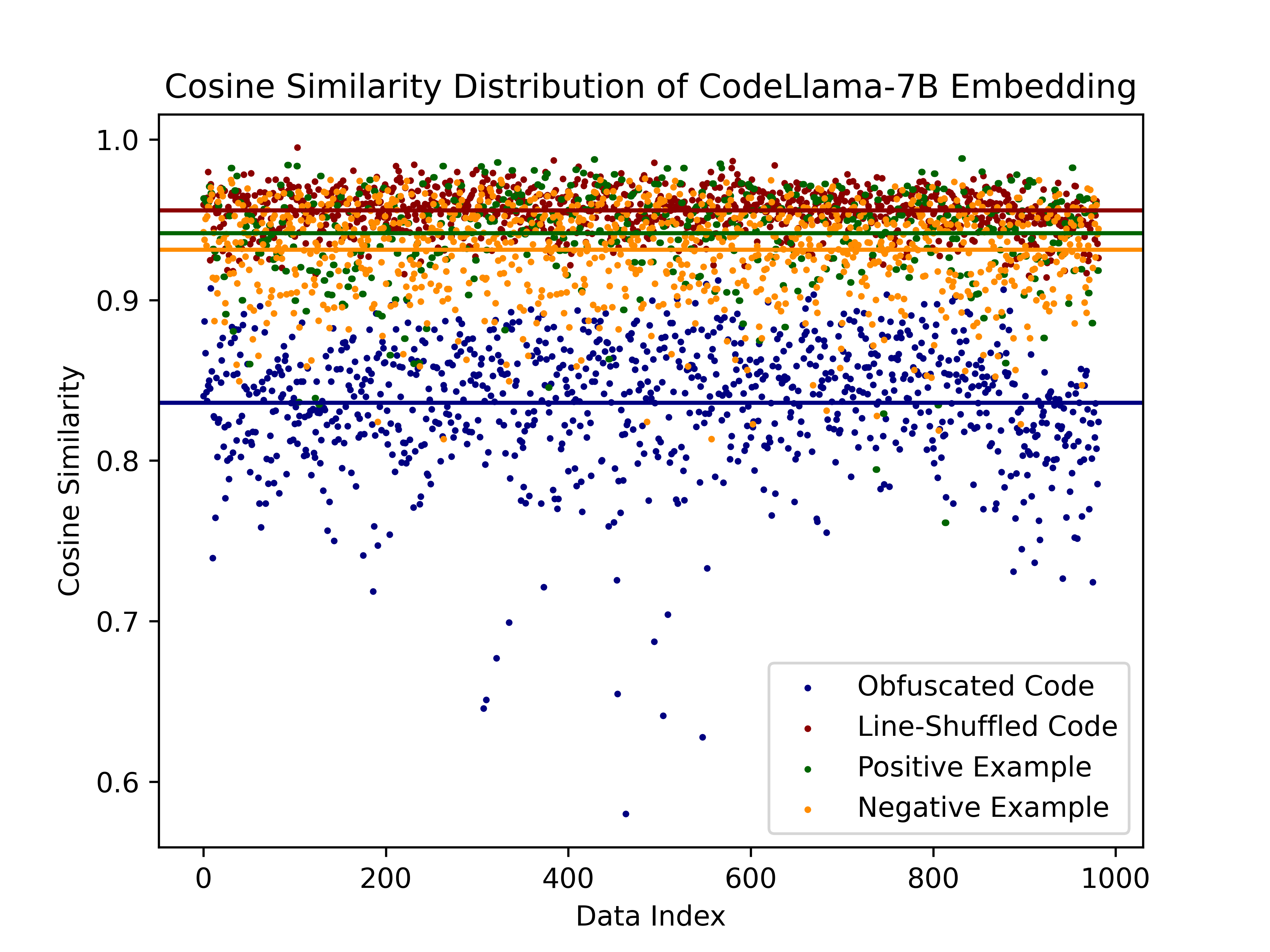}
  \end{minipage}
  \begin{minipage}[b]{0.45\textwidth}
    \includegraphics[width=\textwidth]{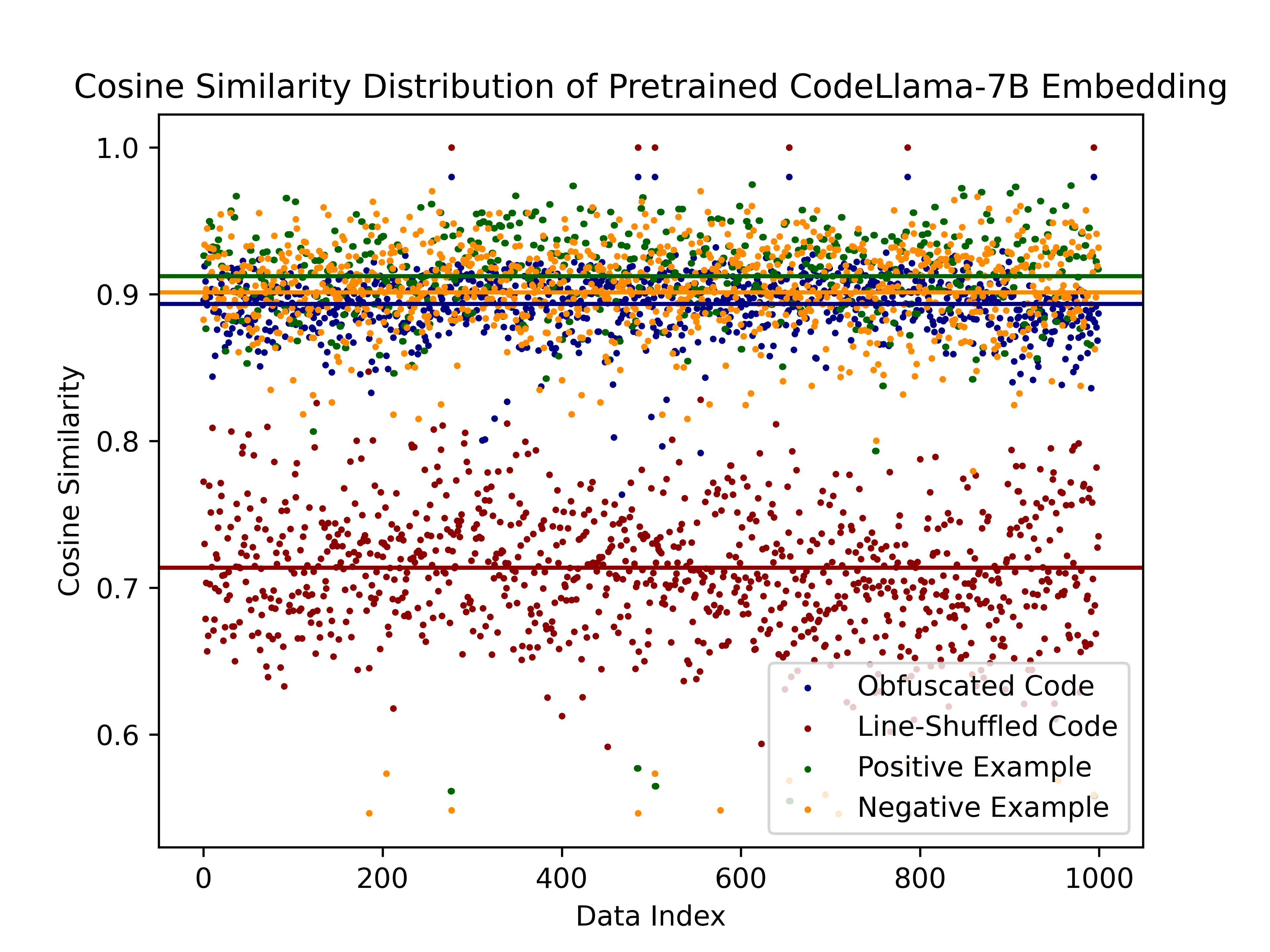}
  \end{minipage}
  \caption{\textbf{Cosine Similarity Visualization.} The left image represents the sentence embedding from Code Llama 7B model, while the right image depicts the result that after our continued pretraining. We differentiate the similarity distance values of various types of codes to the original code with distinct colors. The auxiliary line denotes their average similarity value.}
\label{fig:CodeLlama7b-cosine-similarity}
\end{figure*}

We conducted a deeper analysis on the impact of our proposed pretraining task on the sentence embedding distribution within large language models. We randomly selected 1,000 data pairs from the test dataset of the Code Logic Equivalence Determination task. We calculated the similarity between the obfuscated code, line-shuffled code, positive code, and negative code in relation to the original code, and visualized these similarities in Figure \ref{fig:CodeLlama7b-cosine-similarity}.

We utilized the Code Llama 7B model, both before and after the continuation of pretraining, for comparative experimentation. It was observed that, relative to the original Code Llama 7B model, the average similarity value of the line-shuffled code has significantly decreased, whereas that of the obfuscated code has markedly improved. The distribution of both has returned to a range that aligns with human intuition. Furthermore, we discovered that the similarity of positive and negative codes also decreased following continued pretraining, dropping from approximately 0.95 to around 0.9. This indicates that their distribution within the high-dimensional embedding space has become more dispersed, demonstrating that our method can alleviate the representation degeneration problem existing in LLMs.

\subsection{Ablation Experiment Results}

To investigate the contribution of each component within the pretraining task, we utilize the Code Llama 7B model to carry out experiments on the Code Logic Equivalence Determination task. As depicted in Table \ref{table:abalation study}, the results indicate that conducting the next prediction token prediction experiment solely on the Java and Python corpora can marginally enhance the performance on the D(Positive) > D(Negative) subtask. The performance on the other two tasks, however, does not exhibit any significant changes. The next token prediction task on the hard positive code tends to slightly exacerbate the model's misinterpretation of line shuffled code. Conversely, the hard negative code component does not influence the model's comprehension of obfuscated code.

\captionsetup[table]{font={small}}
\begin{table}[ht]

\begin{adjustbox}{width=\linewidth}
\begin{tabular}{l|ccc}
\hline
\textbf{Model} & $D(Pos)>D(Neg)$ & $D(Obf)>D(Neg)$ & $D(Pos)>D(LSh)$ \\
\hline
Code Llama 7B & {66.91\%} & {1.54\%} & {36.82\%} \\
+ Ori & {75.8\%} & {3.66\%} & {47.15\%} \\
+ Obf & {68.66\%} & {42.2\%} & {32.13\%} \\
+ Lsf & {68.32\%} & {1.83\%} & {72.27\%} \\

\bottomrule
\end{tabular}
\end{adjustbox}

\caption{\raggedright {\textbf{Abalation study of pretraining task in Code Logic Equivalence Determination Task.} "+ Ori" column represents the performance of the next token prediction task only on the original code; in the following two columns, we assess the pretraining part of the obfuscated code and the line shuffled code in turn.} }
\label{table:abalation study}

\end{table}

\section{Conclusion}
In this paper, we highlighted the limitations of current large language models in comprehending code syntax akin to human understanding. Through the implementation of various interference schemas, we have illuminated the discrepancies between these models and human intuition in dealing with functionally equivalent but textually diverse codes and subtly distinct codes that instigate significant bugs. To quantify this discrepancy, we introduced the Code Logic Equivalence Determination task. We also proposed a novel pretraining task that expands on GPT's next token prediction, and experiments have demonstrated that following this pretraining, both Code Llama and StarCoder showed substantial improvement on our new task and the code completion task. This suggests that our approach can effectively enhance the performance of large language models in code-related tasks. The findings of this study open up new avenues for future research to further improve the ability of AI models to understand and generate code akin to human programmers.

\section{Limitations}
In our study, we have only explored the impact of certain perturbations on large language base models in the two programming languages, Java and Python. Looking forward, we aim to introduce more intricate forms of code perturbations, such as logical symbol replacement, return type replacement, among others, to attain a more comprehensive understanding of the current large language models' ability to comprehend code.

\bibliography{anthology,main}
\bibliographystyle{acl_natbib}

\appendix

\section{Extra Perturbations Analysis}
\label{sec:Extra_Perturbations_Analysis}
In this section, we present in this section some extra forms of perturbation that can help us more fully explore the perspective of large language models understanding code.

\noindent \textbf{Token Shuffled Perturbation} It represents a situation with greater disturbance amplitude than line shuffled code. Obviously, the line shuffling will impact the functionality of the code but humans can still briefly guess the original functionality based on the sequential information in each line. However, the token shuffling will absolutely break the original code into pieces and the perturbed code is hardly readable.

\noindent \textbf{Replace-based Perturbations.} Beyond variable names, we are intrigued by the influence of the remaining textual elements of the code on the model's comprehension of the code's overall semantics. We suggest replacing certain tokens from the original code to discern which sections of the code exert a greater influence on the embeddings. There are two types of replace-based perturbations to consider: (1) Replacing reserved keywords (e.g., 'assert', 'const', and 'if') in the code, and (2) Replacing structure-related symbols, including brackets and specific punctuation symbols (i.e., '(', ')', '[', ']', '{', '}', ',', '.', and ';') within the code. The first method of substitution will disrupt the semantic information conveyed by many reserved keywords, while the second method will impact the code's grammatical structures. We collect common keywords and symbols in Java, ensuring that each random selection can choose a token differing from the one in the original position. Note that deleting some punctuation symbols directly may cause their neighboring tokens to be merged into one token, we simply replace the token with a blank space instead to avoid tokenization differences.

We construct them in the same way as section 3. They are all rejected candidates, so they form a test data pair with the positive code. We present the performance of Code Llama and StarCoder on six subtasks as a whole and display it in Figure \ref{fig:Embedding_Similarities_Analysis_all}. 

Upon analyzing Replace-based Perturbations, we find that code with replaced symbols exhibits more pronounced information loss compared to code with replaced keywords. This is evident in the fact that several tested models can readily identify the errors contained within them. In conjunction with the obfuscated code experiment, we can initially infer that when the model interprets the code, it prioritizes variable names, symbols, and keywords, in that order. 

\section{Prompt QA Similarities Analysis}
\label{sec:Prompt_QA_Similarities_Analysis}

Beyond examining the distribution of these codes within the embedding space, we have also approached the model's interpretation of these perturbation codes from an alternative perspective. Simulating the related question-and-answer task, we utilized a purposely designed prompt template shown in blow to scrutinize the model. 

\begin{spverbatim}
SYSTEM MESSAGE:
You are a Java expert who can assess code based on its relevance to the query. 
USER PROMPTS:
Please select the most relevant to the given query code from the following two candidate codes. Use the demo example as a reference to understand the format.

Demo Example:
Query: 
'''java
public void main(int num) { 
    while(num<20) {
         System.out.print(num);
         num++;
      }
}  
'''
Candidate A: 
'''java
public void F(int V) { 
    while(V<20) {
         System.out.print(V);
         V++;
      }
}  
'''
Candidate B: 
'''java
public void main(int num) { 
    while(num<20) {
         num++;
         System.out.print(num);
      }
} 
'''
Answer: A

Your task:
Query:
'''java
{Query Code}
'''
Candidate A: 
'''java
{Candidate A Code}
'''
Candidate B: 
'''java
{Candidate B Code}
'''
Answer:
\end{spverbatim}

We investigated the model's potential for error in the aforementioned seven scenarios by querying which of the two candidate codes it considered to be more closely related to the original code. We employs a template derived from the question-answer task, which provides a vivid demonstration to streamline the model's output format. The final selection comprises model completions that meet the stipulated output format and excludes samples that do not adhere to the anticipated generated form (A or B). Experimental procedures were implemented on the six subtasks proposed in Section 3, the specific accuracies of the responses are delineated in Fig \ref{fig:Prompt QA Similarity Analysis}.

The results from the embedding similarity analysis experiments indicate that Code Llama and StarCoder still struggle with comprehending obfuscated code and shuffled line code. The discrepancy between the model and human intuition becomes particularly evident when dealing with functionally equivalent but textually diverse codes, and slightly variant codes that lead to significant bugs, regardless of whether the similarity distance is calculated from the embedding space or the model is compelled to generate answers through template design.

These experimental observations could further substantiate our hypothesis that large models primarily rely on key tags when analyzing code, leading to a limited understanding of the underlying logic and subsequently poor performance. This outcome is not attributable to the large language model's weak sentence embedding representation, as one might initially assume.

\section{Models for Experiments}
\label{app:Models for all experiments}
The selected models for comparison in our experiments include:
\begin{itemize}
    
    \item \textbf{Code LLaMA} \citep{roziere2023code}. Code Llama is a code-specialized version of Llama2 \citep{touvron-2023-llama2} trained on code-specific datasets. In our experiments, we used the base version model hosted on the Hugging Face\footnote{\url{https://huggingface.co/}} platform.
    \item \textbf{StarCoder} \citep{li-2023-starcoder}. StarCoder is a widely-used large code language model trained on diverse sources, including 80+ programming languages, Git commits, GitHub issues, and Jupyter notebooks. It's also one of the foundation models in our paradigm experiments. In our experiments, we used the base version model hosted on the Hugging Face platform.
    \item \textbf{CodeGPT-adapted} \citep{lu2021codexglue} is proposed in the paper of CodeXGlue benchmark. It begins with the GPT-2 model and continues training on the code corpus, which uses the GPT-2 vocabulary and retains its natural language understanding capabilities. By default, the CodeGPT-adapted model is used for code completion and text-to-code generation tasks.
    \item \textbf{PyCoder} \citep{takerngsaksiri2024syntax} is an automated, syntax-aware code completion tool. It uses token type information for training and doesn't require complete source code. It employs Multi-Task Training techniques for token and type prediction tasks.
    \item \textbf{CodeT5+} \citep{wang-2023-codet5+}. CodeT5+ is a new family of open code LLMs with an encoder-decoder architecture trained on various pretraining tasks. InstructionCodeT5+ is further fine-tuned on the Code Alpaca dataset.
    \item \textbf{gpt-3.5-turbo} \citep{chatgpt} originated from ChatGPT, it is similar to the training method of InstructGPT, and adding more high-quality training data, making it be the currently one of the most competitive code generation models.
    \item \textbf{gpt-4} \citep{openai-2023-gpt4} is the top large-scale language model currently open to OpenAI. It has achieved the best results in many evaluation tasks. Here we use the gpt-4-0613 version.
    
\begin{figure*}[!htb]
\center{\includegraphics[width=15cm]  {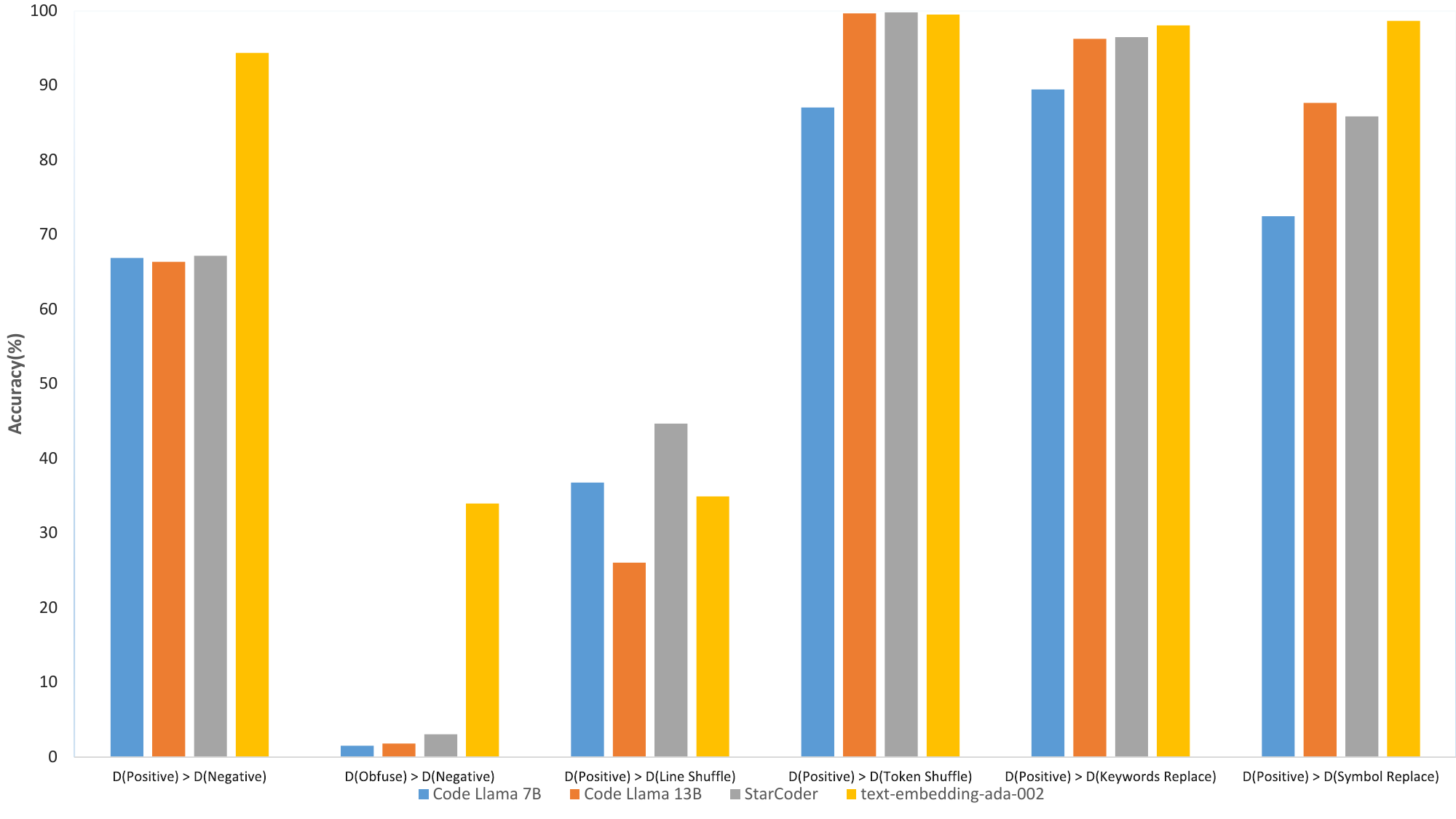}}
\caption{\textbf{The accuracy distribution of different models on different embedding similarity analysis subtasks.} The blue, orange, and gray columns respectively represent the accuracy of large language models of different sizes (7B, 13B, 15B) on the six distance analysis tasks we have designed. Additionally, we have included OpenAI's specialized embedding model, text-embedding-ada-002 (depicted by a yellow column), as a control for comparison.}
\label{fig:Embedding_Similarities_Analysis_all}
\end{figure*}

\begin{figure*}[!htb]
\center{\includegraphics[width=15cm]  {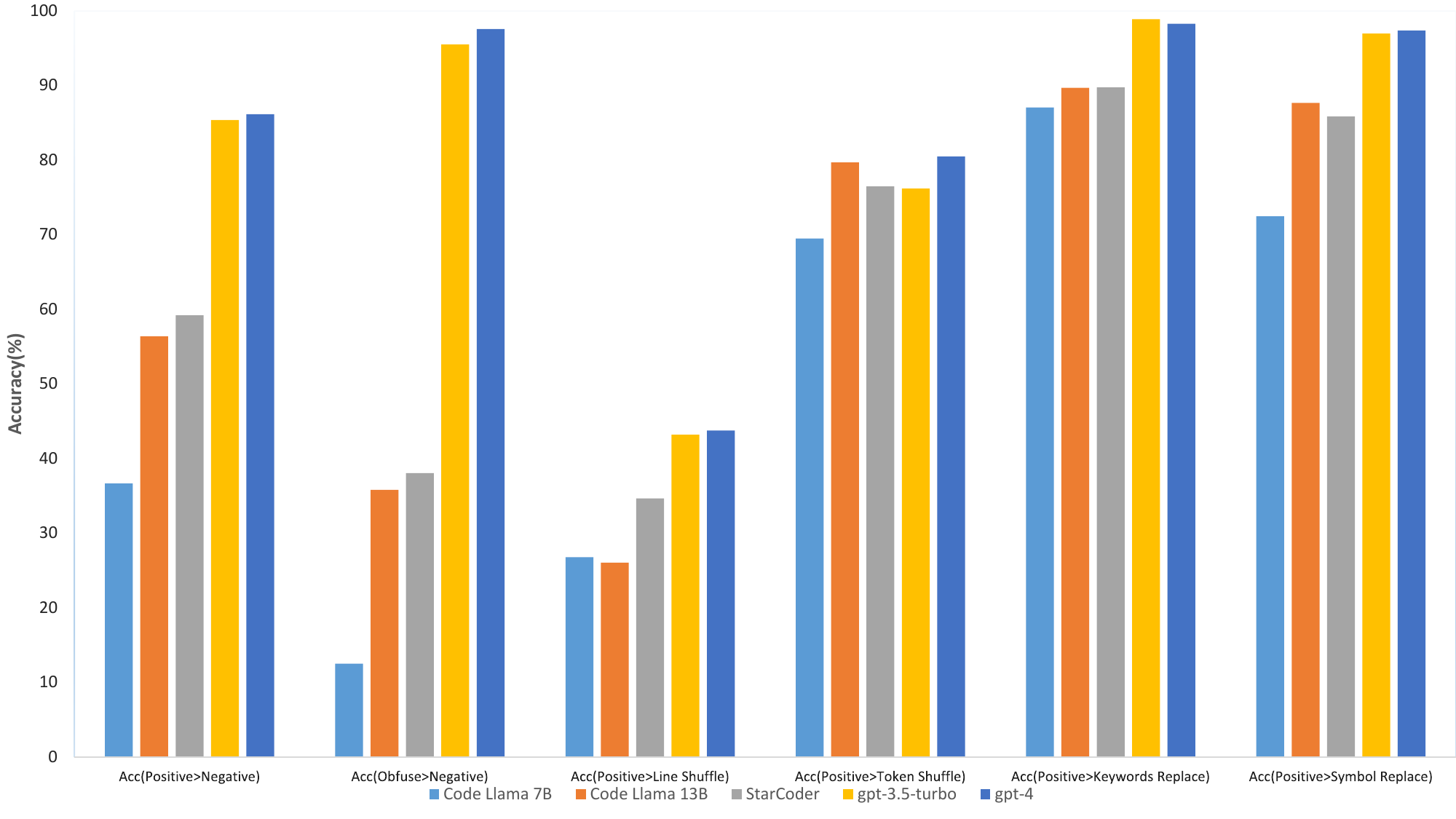}}
\caption{\textbf{The accuracy distribution of different models on different prompt question-answer similarities analysis subtasks.} The term $Acc(Positive>Negative)$ signifies the accuracy of the model in selecting the appropriate option of positive code when presented with two candidates, namely, the positive and negative codes. The remaining cases can be inferred similarly. We have selected OpenAI’s gpt-3.5-turobo and gpt-4 models as benchmarks for this analysis.}
\label{fig:Prompt QA Similarity Analysis}
\end{figure*}

\section{Training Details}
\label{sec:training_details}

For both foundation models, we conduct training on Azure Machine Learning Studio's cluster \footnote{\url{https://ml.azure.com/}}, utilizing 4 nodes, each equipped with 8 V100 GPUs featuring DeepSpeed Zero-3 \citep{rajbhandari-2019-zero} offload. In both the pre-training phase and the fine-tuning phase, we use full parameter training and enable fp16 accuracy. In the code generation task, we use the same training parameters (learning rate as 2e-5 and epoch as 2) for the original model and the model after "next token prediction+" pre-training.

\section{Specific Examples of Perturbations}
\label{sec:Specific_Examples_of_Perturbations}
In this section, we take the solution of a Java problem of "Two Sum" as an example to show the specific situation after adding different perturbations. The figure below shows all the specific examples of perturbations.

\begin{figure}[!htb]
\center{\includegraphics[width=7.5cm]  {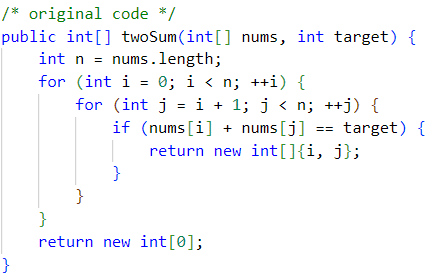}}
\caption{\textbf{The example of original code.} }
\label{fig:origin code}
\end{figure}

\begin{figure}[!htb]
\center{\includegraphics[width=7.5cm]  {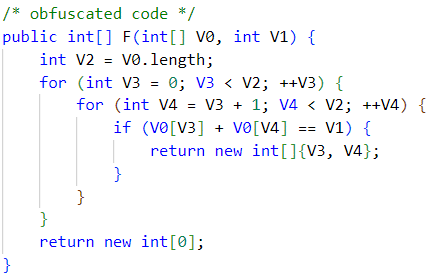}}
\caption{\textbf{The example of obfuscated code.} }
\label{fig:obfuscated code}
\end{figure}

\begin{figure}[!htb]
\center{\includegraphics[width=7.5cm]  {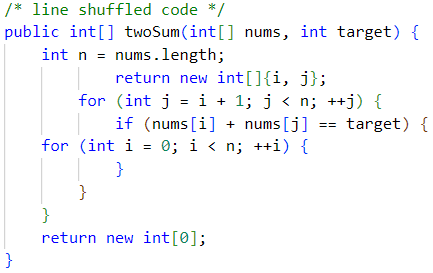}}
\caption{\textbf{The example of line shuffled code.} }
\label{fig:line shuffled code}
\end{figure}

\begin{figure}[!htb]
\center{\includegraphics[width=7.5cm]  {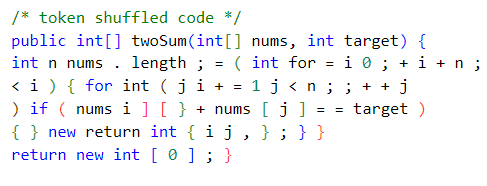}}
\caption{\textbf{The example of token shuffled code.} }
\label{fig:token shuffled code}
\end{figure}

\begin{figure}[!htb]
\center{\includegraphics[width=7.5cm]  {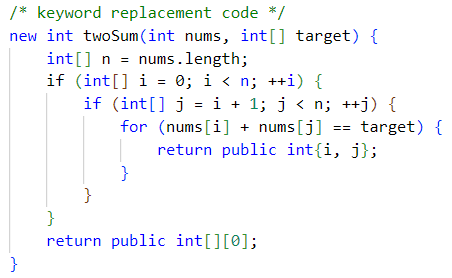}}
\caption{\textbf{The example of keywords replacement code.} }
\label{fig:keywords_replacement_code}
\end{figure}

\begin{figure}[!htb]
\center{\includegraphics[width=7.5cm]  {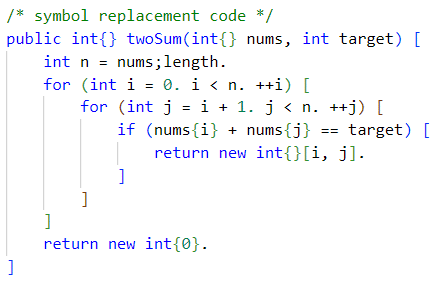}}
\caption{\textbf{The example of symbol replacement code.} }
\label{fig:symbol replacement code}
\end{figure}
 
\end{itemize}

\end{document}